\documentclass[]{spie}  %>>> use for US letter paper
\usepackage{amsmath,amsfonts,amssymb}
\usepackage{graphicx}
\usepackage[colorlinks=true, allcolors=blue]{hyperref}

\title{Motion correction of dynamic contrast enhanced MRI of the liver}

\author[1]{Mari\"elle J.A. Jansen}
\author[2]{Wouter B. Veldhuis}
\author[2]{Maarten S. van Leeuwen}
\author[1]{Josien P.W. Pluim}
\affil[1]{Center for Image Sciences, University Medical Center Utrecht, the Netherlands}
\affil[2]{Department of Radiology, University Medical Center Utrecht, the Netherlands}

\authorinfo{Corresponding author: M.J.A.Jansen-35@umcutrecht.nl}

\begin{document} 
\maketitle

\begin{abstract}
Motion correction of dynamic contrast enhanced magnetic resonance images (DCE-MRI) is a challenging task, due to changes in image appearance. In this study a groupwise registration, using a principle component analysis (PCA) based metric,\cite{Huizinga2016} is evaluated for clinical DCE MRI of the liver. The groupwise registration transforms the images to a common space, rather than to a reference volume as conventional pairwise methods do, and computes the similarity metric on all volumes simultaneously. \\
\\
This groupwise registration method is compared to a pairwise approach using a mutual information metric. Clinical DCE MRI of the abdomen of eight patients were included. Per patient one lesion in the liver was manually segmented in all temporal images (N=16). The registered images were compared for accuracy, spatial and temporal smoothness after transformation, and lesion volume change. Compared to a pairwise method or no registration, groupwise registration provided better alignment.\\
\\
In our recently started clinical study groupwise registered clinical DCE MRI of the abdomen of nine patients were scored by three radiologists. Groupwise registration increased the assessed quality of alignment. The gain in reading time for the radiologist was estimated to vary from no difference to almost a minute. A slight increase in reader confidence was also observed. Registration had no added value for images with little motion. \\
\\
In conclusion, the groupwise registration of DCE MR images results in better alignment than achieved by pairwise registration, which is beneficial for clinical assessment.
\end{abstract}

% Include a list of keywords after the abstract 

\section{INTRODUCTION}
\label{sec:intro}  % \label{} allows reference to this section

Dynamic contrast enhanced MRI (DCE-MRI) is an important part of the MR examination of the liver. DCE-MRI provides insight in the microcirculation and tissue characteristics of liver lesions and parenchyma.\cite{Choyke2003} During the acquisition of DCE-MRI, a bolus of contrast agent (CA) is intravenously injected. The distribution of the CA is followed by repeatedly imaging the organ of interest. In the clinical routine, data acquisition takes several minutes and to limit breathing motion, the patient is asked to hold his/her breath several times.\par
Unfortunately the resulting images can still suffer from motion caused by inconsistent breath-hold depth or gasping for air, as well as by cardiac or bowel movement.\cite{Wollny2012} Patient motion complicates analysis of time series. Motion correction of DCE-MRI time series allows more accurate analysis of the tissue characteristics, easier assessment by radiologists and it also reduces motion artefacts in subtraction images.\par
Correcting for motion requires registration of the time series. Conventional pairwise registration methods generally do not achieve the temporal smoothness needed for visual inspection, due to the difficulty of separating the intensity changes caused by motion and contrast. Therefore groupwise registration methods for DCE-MRI have been proposed, to distinguish this motion from intensity changes due to contrast using models or data reduction techniques.\cite{Buonaccorsi2007,Hayton1997,Bhushan2011,Wollny2012,Melbourne2007,Hamy2014} The disadvantage of using a model, as in Ref.~\citenum{Buonaccorsi2007,Hayton1997,Bhushan2011}, is that a model could be corrupted by noise and acquisition artefacts, or it might not represent the data set, leading to inaccurate or inappropriate model fits.\textsuperscript{1,4} The disadvantage of data reduction techniques is that registration towards the created synthetic images, as in Ref.~\citenum{Wollny2012,Melbourne2007,Hamy2014}, can cause a bias towards these images. Therefore, a true groupwise method, using neither reference images or a model, such as the method which has been proposed by Huizinga et al. (2016), is preferred\cite{Huizinga2016}. This method has been validated on several different data sets, but not yet on clinical data of the DCE-MRI of the abdomen with a limited number of temporal images.\par
This study evaluated the generic groupwise registration method proposed by Huizinga et al. (2016) for motion correction of DCE-MRI images of the liver. The method was compared with a conventional pairwise approach, using the mutual information (MI) metric. In addition, we investigated the clinical benefit of groupwise registration of DCE-MRI in lesion assessment.

\section{METHODS}
\subsection{Data}
\label{sec:Data}

Data of patients from the University Medical Center Utrecht were included. The data were acquired on a 1.5T MR scanner (Philips Medical Systems) using a clinical protocol with the following parameters: TE: 2.143 ms; TR: 4.524 ms; flip angle: 10 degrees. DCE-MRI data were acquired in several breath holds and during each breath hold 1 to 5 images were taken. After the first image, gadobutrol (0.1 ml/kg Gadovist of 1.0 mmol/ml at 1 ml/s) or gadoxetate disodium (0.1 ml/kg Primovist of 0.25 mmol/ml at 1 ml/s) was administered at once, followed by 25 ml saline solution at 1 ml/s. In total 16 3D images per patient were acquired with 100-105 slices and a matrix of 256$\times$256. Voxel size was 1.543 mm $\times$ 1.543 mm $\times$ 2 mm. \par
Eight data sets were registered using the pairwise and the groupwise registration and were used for comparison of the registration approaches. In these eight data sets one visible lesion was manually annotated by the researcher in all the sixteen images of the DCE-MRI series. In some images, the lesion was hard to distinguish from the liver parenchyma, but lesion location and shape in other images, in which the lesion was clearly visible due to the contrast agent uptake, were used as prior information. The lesions in three data sets were annotated twice and the intra-observer Dice Similarity Coefficient (DSC) was 0.81.\par
Nine data sets were used in the clinical evaluation. For six data sets, the clinical focus was on the liver; for two on the pancreas and for one the kidneys. The data was only registered by the groupwise approach, limiting the clinical evaluation to the best registration method, according to the comparison of registration approaches. After registration, subtraction images were obtained by subtracting the first, non-contrast image from the images after contrast injection. The transformed data sets and their corresponding subtraction images were sent to PACS, where they can be used for routine evaluation of patient scans.

\subsection{Registration methods}
All data was registered using Elastix\cite{Elastix} for both the groupwise approach and pairwise approach.

\subsubsection{Groupwise registration method}
The groupwise registration method registers all images simultaneously to a common space, taking into account the intensity change over time due to contrast. The method feeds a dissimilarity metric using principle component analysis on a correlation matrix.\textsuperscript{1} The correlation matrix C is defined as: $C = \frac{1}{N-1}  S^{-1} (M - \bar{M})^T (M - \bar{M}) S^{-1}$  , where N is the number of voxels per 3D volume, M is the $N \times V$ matrix, and $V$ is the number of volumes in the DCE-MRI series, $\bar{M}$ ̅is a matrix where each element in the column is the column-wise mean of $M$, and $S$ is a diagonal matrix with the column-wise standard deviations of $M$. The dissimilarity metric is defined as: $D_{PCA} = \sum_{j=1}^{V} j\lambda_j $, where $\lambda_j$ is the $j$\textsuperscript{th} eigenvalue, obtained by applying PCA to the correlation matrix $C$. Each eigenvalue is inversely proportional weighted, i.e. the first and highest eigenvalue is weighted once and the last and lowest eigenvalue is weighted 16 times in this study. In the registration method, the highest eigenvalues of the correlation matrix are maximised, and so impose more alignment by minimizing the dissimilarity metric.\\

\subsubsection{Pairwise registration method}
The pairwise registration method used the conventional mutual information metric. The first, non-contrast image in a series was used as the reference to which all other images in the series were registered. \\
\\
Both pairwise and groupwise method applied a third order b-spline interpolation, had 500 iterations per resolution and selected 2048 samples randomly during each iteration. The final grid spacing was set to 16 mm. The number of resolutions was different; three for the pairwise approach and four for the groupwise approach. The settings were determined empirically. 

\begin{figure} [ht]
	\begin{center}
		\begin{tabular}{c} 
			\includegraphics[width=17cm]{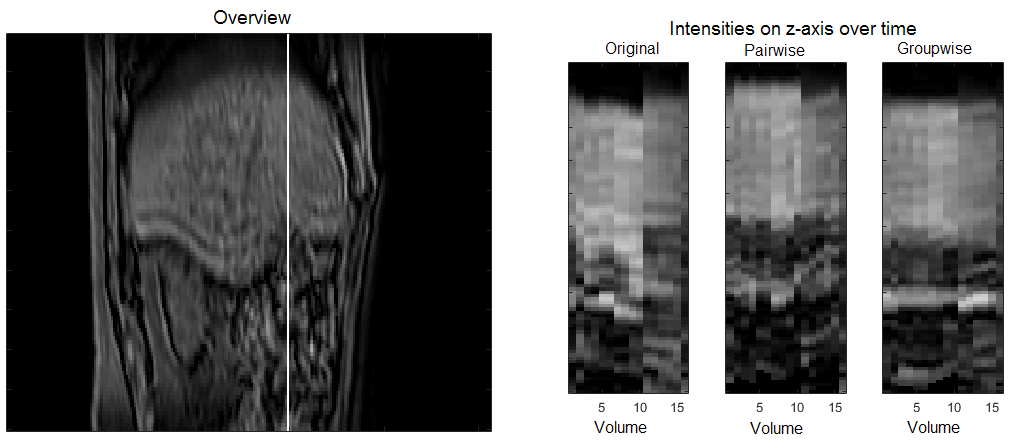}
		\end{tabular}
	\end{center}
	\caption[example]
	{ \label{fig:Overview} 
		Registration result for one patient. Left: slice indicating a single column through the liver (white line). Right: intensities of the indicated line over the 16 images unregistered and two variants of registered images.}
\end{figure} 

\subsection{Experiments}
\subsubsection{Comparison of registration approaches}
The ability of the registration methods to align the images was assessed by calculating the DSC between the lesion volume mask of the first, non-contrast, image and the (transformed) lesion volumes of the other images in the DCE-MRI series for the original (unregistered), pairwise and groupwise registered series. In case of the groupwise approach the lesion volume mask of the first image was transformed from the reference space to the image space according to the registration, because the groupwise method also transforms the first (reference) image to the common space. \par
In addition, the groupwise DSC was calculated to assess the accuracy of the temporal alignment of all lesions in the series. Groupwise DSC is defined as: $16 \times \frac{S_1 \bigcap S_2 ... \bigcap S_{16}}{S_1 + S_2 ... + S_{16}}$, where $S_n$ is the $n^{th}$ manual lesion segmentation. This value will probably not be as high as the pairwise DSC, because a misalignment in one image has a large influence on the groupwise DSC. \par
The smoothness of the intensities over time is another measure of the accuracy of registration methods. For this, the lesion volume of the first non-contrast image was propagated to the other images and for each image the mean intensity in the propagated region was calculated. For the groupwise method, the lesion volume mask was first transformed to the non-contrast image of the groupwise result, before being applied to all the images. The standard deviation (SD) of the second derivative of these mean intensities was calculated to obtain a measure for the smoothness of the intensities over time. A lower SD of the second derivative suggests better alignment.\par
The Jacobian determinant of the transformation defines the volume expansion or shrinkage for each voxel. The mean Jacobian determinant in the lesion volume was calculated for each image to identify volume changes. A mean of 1 is expected since large volume changes would be unrealistic and therefore introduced by the transformation.\par
The SD of the Jacobian determinant is a measure of the smoothness of the transformation in the spatial domain. A small SD implies a smooth spatial transformation without unrealistic deformations. The SD of the Jacobian determinant in the liver region was obtained for each image, except the first, non-contrast image. The mean standard deviation of the 15 volumes was calculated for each DCE MR series.

\subsubsection{Clinical evaluation}
In the clinical study, three radiologists were asked to evaluate and score the groupwise registered images on four points during routine evaluation of patient scans. The quality of the alignment in the images before and after registration was graded by the radiologists. The subtraction images were assessed on the quality and usefulness of the images. The possible time gain or loss reading the registered DCE-MRI series and its subtraction in comparison to the original series was estimated. The reader confidence, when using the registered images and their subtraction in comparison with the original images, was determined by the radiologists as equal, increased or decreased. An overview of the four points and the corresponding meaning of the scores is given in Table \ref{tab:Overview}.\par

\begin{table}[ht]
	\caption{Overview of the four evaluation points and the corresponding scores.} 
	\label{tab:Overview}
	\centering
\begin{tabular}{ccccc}
	\hline
	\textbf{Score} & \textbf{Quality of alignment}                                                                                            & \textbf{\begin{tabular}[c]{@{}c@{}}Quality\\ subtraction\\   images\end{tabular}}      & \textbf{Estimated time}                                                                   & \textbf{\begin{tabular}[c]{@{}c@{}}Reader \\ confidence\end{tabular}} \\ \hline
	1              & \begin{tabular}[c]{@{}c@{}}Poor and assessment\\   is (nearly) impossible\end{tabular}                                   & \begin{tabular}[c]{@{}c@{}}Poor with no\\ diagnostic value\end{tabular}                & \begin{tabular}[c]{@{}c@{}}Time loss of $\geq$ \\   1 minute\end{tabular}                       & Decreased                                                             \\ \hline
	2              & \begin{tabular}[c]{@{}c@{}}Moderate with motion \\ visible\end{tabular}                                                  & \begin{tabular}[c]{@{}c@{}}Moderate with little \\ diagnostic value\end{tabular}       & \begin{tabular}[c]{@{}c@{}}Time loss of almost \\ a minute\end{tabular}                   & Equal                                                                 \\ \hline
	3              & \begin{tabular}[c]{@{}c@{}}Good with hardly any\\   motion in lesion, but\\   some motion in rest of liver\end{tabular}                                        & \begin{tabular}[c]{@{}c@{}}Good with \\ diagnostic value, \\ but not used\end{tabular} & \begin{tabular}[c]{@{}c@{}}Approximately the\\   same time ($\pm$ 10 \\ seconds)\end{tabular} & Increased                                                             \\ \hline
	4              & \begin{tabular}[c]{@{}c@{}}Good with no motion in \\ lesion, but at edges of liver \\ some motion remaining\end{tabular} & \begin{tabular}[c]{@{}c@{}}Good and used for\\   diagnosis\end{tabular}                & \begin{tabular}[c]{@{}c@{}}Time gain of almost \\ a minute\end{tabular}                   & NA                                                                    \\ \hline
	5              & \begin{tabular}[c]{@{}c@{}}Nearly perfect alignment of \\ the lesion and liver/kidney \\ structures\end{tabular}         & \begin{tabular}[c]{@{}c@{}}Good and \\ determinative for \\ diagnosis\end{tabular}     & \begin{tabular}[c]{@{}c@{}}Time gain of $\geq$ 1\\   minute\end{tabular}                       & NA                                                                    \\ \hline
\end{tabular}
\end{table}

\section{RESULTS}
Figure \ref{fig:Overview} shows the result of registration for one patient. The three images on the right-hand side show the intensities of a single line through a 3D volume over the 16 volumes. This line is indicated in the image on the left-hand side. The movement in the z-direction, visible along the images, is reduced by both registration methods. The groupwise registration method provides a smoother transition in the time domain than the pairwise registration method.\par 

\subsection{Comparison of registration approaches}
\begin{figure} [ht]
	\begin{center}
		\begin{tabular}{c} 
			\includegraphics[height=10cm]{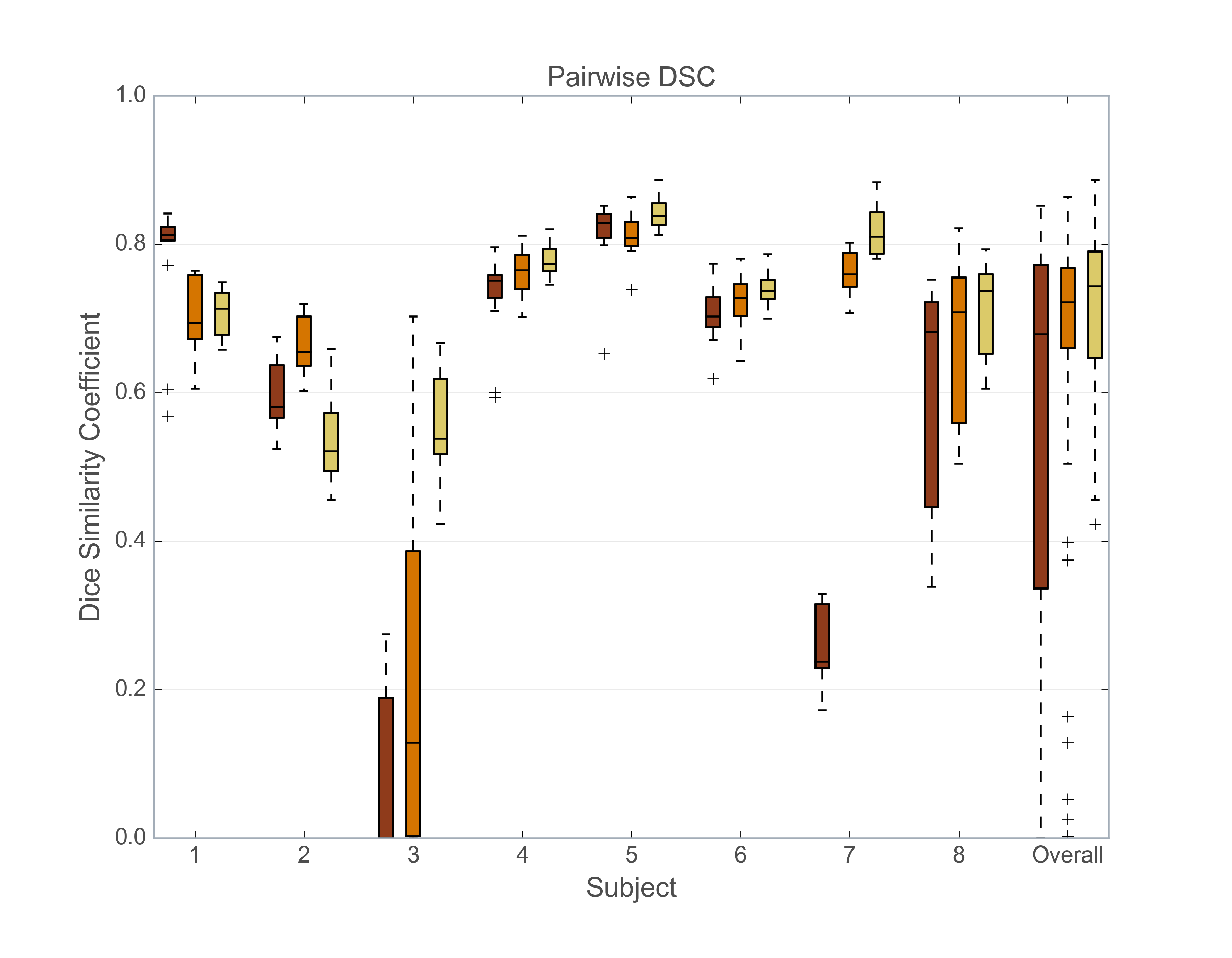}
		\end{tabular}
	\end{center}
	\caption[example]
	{ \label{fig:PairwiseDSC} 
		Boxplot of the DSC between the lesion segmentation of the first image and the other lesion segmentations per subject and the overall result.}
\end{figure}

In Figure \ref{fig:PairwiseDSC} the results of the DSC between the lesion volume of the first, non-contrast image and the 15 (transformed) lesion volumes of the other images in the DCE-MRI series are given in a boxplot per subject and the all subjects together. On average the DSC is 0.57 for the original series, 0.67 for pairwise registration and 0.71 for the groupwise registration. In most cases the DSC increases after registration, especially for the groupwise registration. \par

\begin{figure} [ht]
	\begin{center}
		\begin{tabular}{c} 
			\includegraphics[width=17cm]{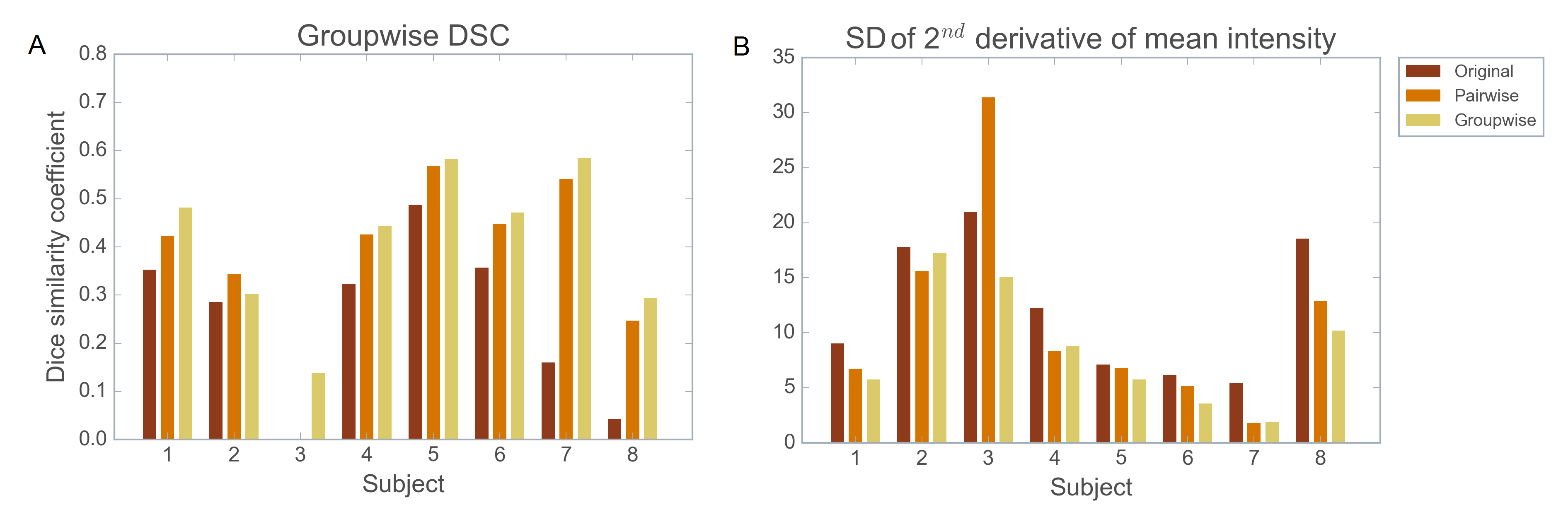}
		\end{tabular}
	\end{center}
	\caption[example]
	{ \label{fig:GroupwiseDSC} 
		A) Groupwise DSC of the lesion per subject. B) SD of the second derivative of the mean intensity in the lesion volume over time per subject.}
\end{figure}

The results of the groupwise DSC of the lesion volumes per patient are shown in Figure \ref{fig:GroupwiseDSC}A. The average groupwise DSC is 0.25 for the original, unregistered images, 0.37 for sequential registration and 0.41 for the groupwise registration. In all cases the groupwise DSC increases after registration. Groupwise registration generally performs better than pairwise registration.\par

The average SD of the second derivative of the mean intensity over time is 12.2 for the original images, 11.1 for the sequentially registered images and 8.5 for the groupwise registered images, suggesting smoother temporal intensity transitions by groupwise registered images. See Figure \ref{fig:GroupwiseDSC}B for the results per subject.\par

\begin{figure} [ht]
	\begin{center}
		\begin{tabular}{c} 
			\includegraphics[height=8cm]{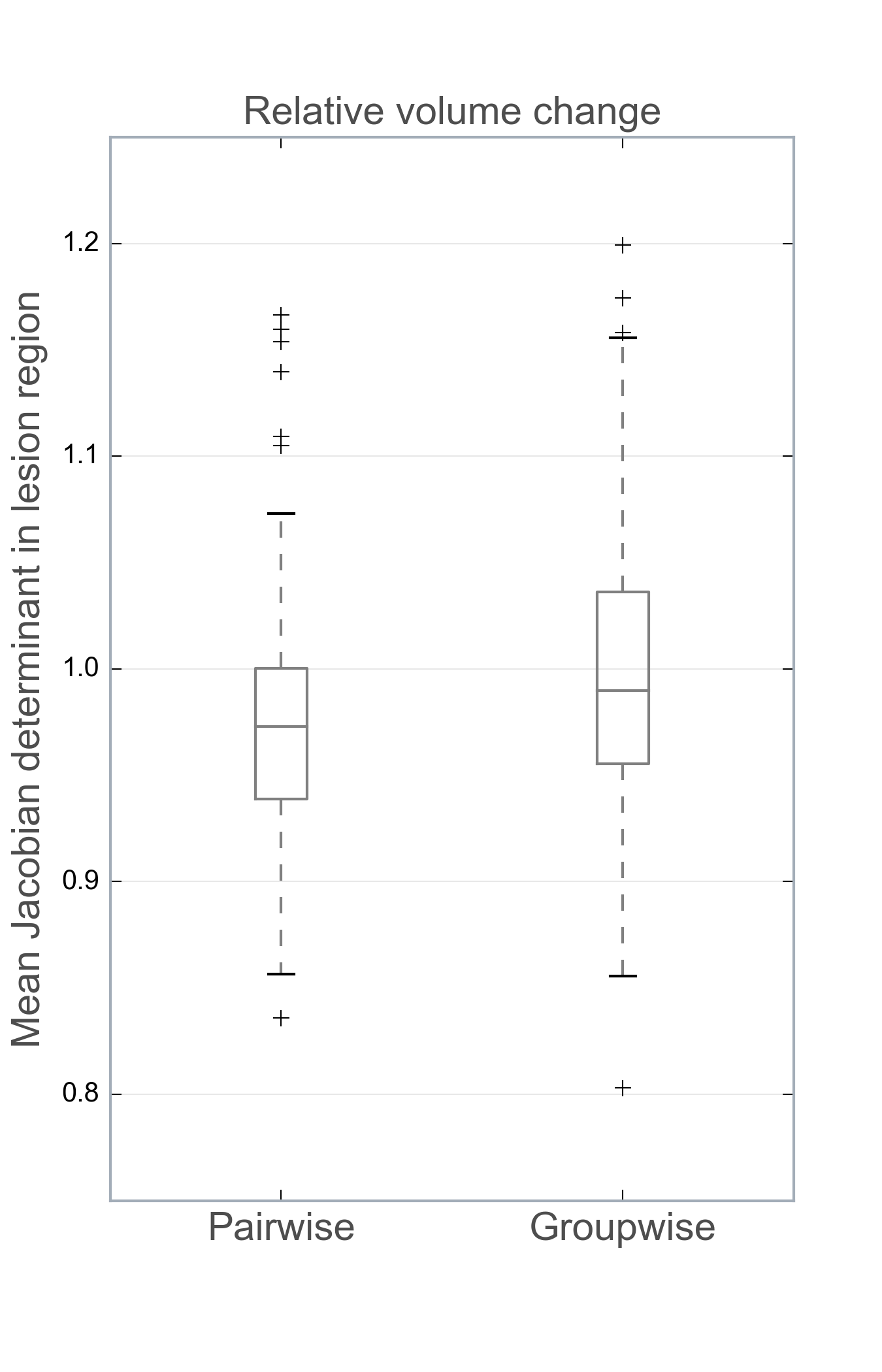}
		\end{tabular}
	\end{center}
	\caption[example]
	{ \label{fig:VolumeChange} 
		Boxplot of the mean Jacobian determinant per lesion volume.}
\end{figure}

The mean Jacobian determinant of all the lesions in all images is 0.974 $\pm$ 0.058 for the pairwise registration and 1.000 $\pm$ 0.067 for the groupwise registration. A Jacobian determinant of 1 would indicate no volume change within the lesion. The lesion volume is kept quite stable in both registration methods. See Figure \ref{fig:VolumeChange} for the boxplot of the mean Jacobian determinant per lesion volume for pairwise and groupwise registration.\par 

The overall mean (range) SD of the Jacobian determinant in the liver region is 0.067 (0.046 - 0.103) for the sequential registration and 0.047 (0.039 - 0.060) for the groupwise registration. This suggests that the latter has a smoother spatial transformation with less extreme deformations. \par 
In general the groupwise registration has the best results. There are two exceptions (subject 1 and 2). The original series of the first subject is nicely aligned in the first four breath holds (i.e. the first 12 images) and after that more motion is present. Therefore the pairwise DSC of the original series in this subject has a higher median than the pairwise DSC of the registration methods. The registration methods have an overall better alignment (see groupwise DSC), but at the cost of pairwise alignment with the first image.\par 
For the second patient the groupwise method failed to align the first non-contrast image with the rest of the images. For this subject the groupwise registration method results of the pairwise DSC and the smoothness of temporal intensity transitions are worse than the pairwise method, because these measures rely on the lesion volume of the first image. Apart from the first image the alignment of the groupwise registration is visually better than the pairwise registration. This is supported by the fact that the mean SD of the Jacobian determinant in the liver region is lower for the groupwise method (0.060) than the pairwise method (0.073), indicating smoother spatial deformations for the former method.

\subsection{Clinical evaluation}
The clinical study is ongoing, but the early results suggest that usage of registered images and their subtraction is beneficial. The average quality of the alignment increases from 2.1 for the original images to 4 for the groupwise registered images. See Figure \ref{fig:VolumeChange} for the scores before and after the groupwise registration. All the original images had a score of 2 or 3, except in one case a score of 5 was assigned. After registration all scores went up or stayed equal.\par 
\begin{figure} [ht]
	\begin{center}
		\begin{tabular}{c} 
			\includegraphics[height=7cm]{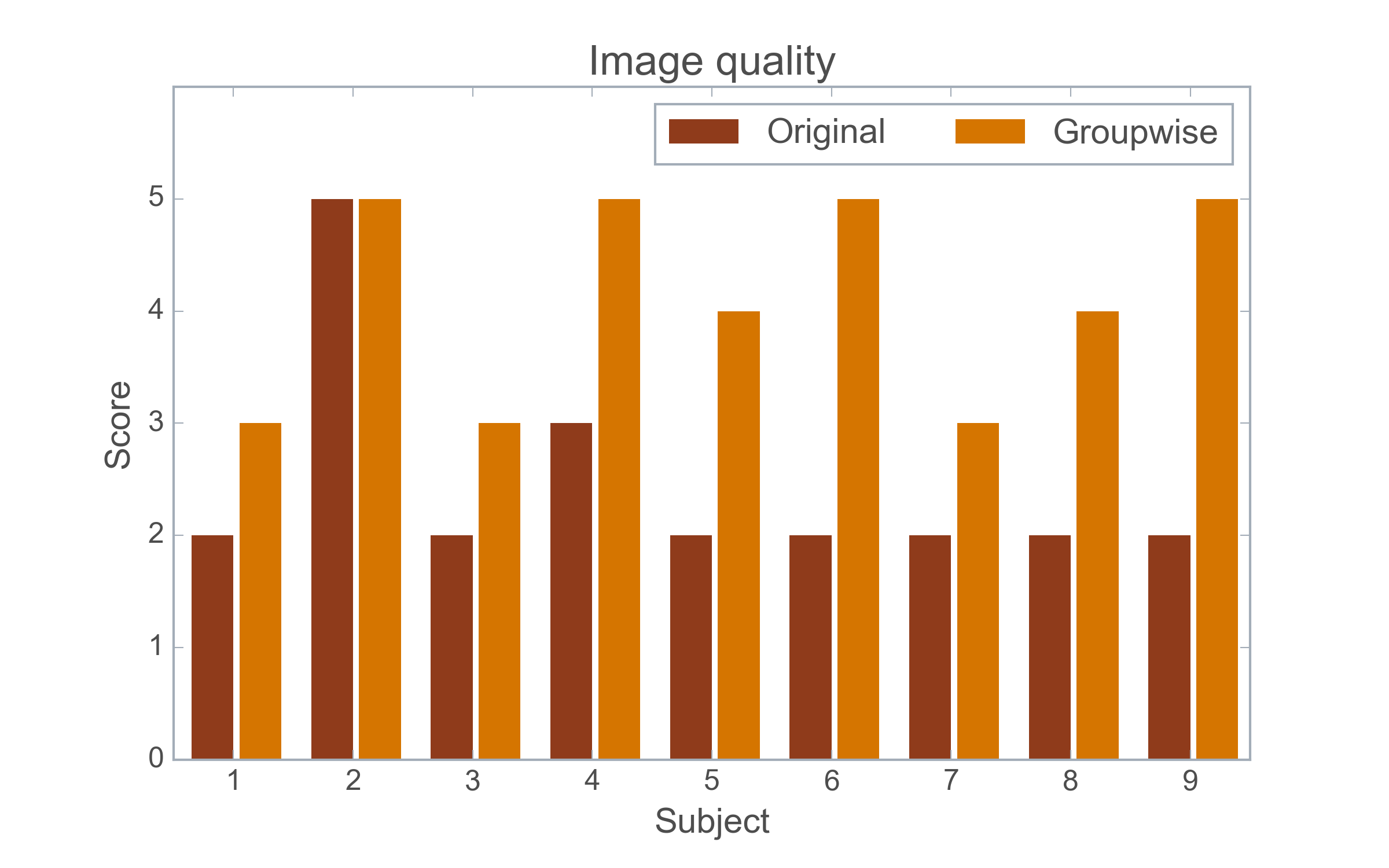}
		\end{tabular}
	\end{center}
	\caption[example]
	{ \label{fig:ImageQuality} 
	Image quality before and after groupwise registration, as judged by radiologists.}
\end{figure}
In all cases the subtraction images were good and used during the routine assessment. In two of them the images were even determinative for clinical assessment. So the subtraction images have an added value in all cases, even when little motion was present in the original series.
For five cases, no difference in reading time was estimated by the radiologists. For the remaining four cases a time gain of less than a minute was estimated when reading the registered images rather than the original images.\par 
Reader confidence stayed equal in six cases and increased for the three others when reading the registered series and their subtraction.

\section{DISCUSSION AND CONCLUSIONS}
\subsection{Comparison of registration approaches}
In this study a comparison was made between pairwise and groupwise registration. A groupwise registration method was shown to provide better alignment of DCE-MRI time series than either a pairwise method or no registration. This is most likely explained by the elimination of bias towards a reference image. Bias was removed through simultaneous registration of the images to the common space and applying the metric to all images at once, instead of separately registering toward a reference image. The outcomes of this study are in agreement with earlier studies using different registration methods and/or data. \cite{Huizinga2016,Melbourne2007,Hamy2014}\par 
Major imaging artefacts, as seen in subject 3, can cause motion correction to fail. Failure of registration could lead to changes in lesion appearance and volume.\cite{Sundarakumar2015} The most extreme volume changes were observed in subject 3 for both the pairwise and groupwise registration method. However, the groupwise registration seems to be more robust against artefacts than the pairwise method, because the pairwise and groupwise DSC results are higher and the SD of the 2\textsuperscript{nd} derivative of the mean intensity is lower for the groupwise registration method.

\subsection{Clinical evaluation}
The clinical study although as yet only performed with nine data sets, already shows the benefits of using registration and supplying a subtraction image. This improved the quality of the DCE-MRI series in terms of motion in all subjects. The preference for registered image series over the original was shown in an earlier study.\cite{Melbourne2007} Besides better visual assessment, the registration also gives the opportunity of supplying good subtraction images. Both the registered images and their subtraction can contribute to increasing reader confidence.\par 
In conclusion, this study shows that groupwise registration using a PCA-based dissimilarity metric on DCE MRI of the abdomen can achieve good and temporally smooth alignment. It achieves better results than pairwise registration with a mutual information metric. Early results from the clinical validation of the groupwise registration show that radiologists benefit from the registration regarding reader confidence while assessing the DCE MR series.

\acknowledgments % equivalent to \section*{ACKNOWLEDGMENTS}    
This work was financially supported by the project BENEFIT (Better Effectiveness aNd Efficiency by measuring and modelling of Interventional Therapy) in the framework of the EU research programme ITEA (Information Technology for European Advancement).

% References
\bibliography{report} % bibliography data in report.bib
\bibliographystyle{spiebib} % makes bibtex use spiebib.bst

\end{document}